%% file: conference_101719.tex
\documentclass[conference]{IEEEtran}
\IEEEoverridecommandlockouts
\usepackage{amsmath,amssymb,amsfonts}
\usepackage{algorithmic}
\usepackage{graphicx}
\usepackage{textcomp}
\usepackage{xcolor}
\usepackage{booktabs}
\usepackage{url}
\usepackage[backend=biber, style=ieee, locallabelwidth]{biblatex}
\bibliography{bibliography.bib}
\newif\ifanonymous
\newif\ifcameraready

\anonymousfalse

\camerareadytrue
\input{aux_commands}
\def\BibTeX{{\rm B\kern-.05em{\sc i\kern-.025em b}\kern-.08em
    T\kern-.1667em\lower.7ex\hbox{E}\kern-.125emX}}
\begin{document}

\title{MLOps: A Multiple Case Study in Industry 4.0}


\author{\IEEEauthorblockN{1\textsuperscript{st} Leonhard Faubel}
\IEEEauthorblockA{\textit{University of Hildesheim} \\
Hildesheim, Germany \\
faubel@uni-hildesheim.de}
\and
\IEEEauthorblockN{2\textsuperscript{nd} Klaus Schmid}
\IEEEauthorblockA{\textit{Univeristy of Hildesheim} \\
Hildesheim, Germany \\
schmid@uni-hildesheim.de}
}

\maketitle

\begin{abstract}

    As Machine Learning (ML) becomes more prevalent in Industry 4.0, there is a growing need to understand how systematic approaches to bringing ML into production can be practically implemented in industrial environments. Here, MLOps comes into play. MLOps refers to the processes, tools, and organizational structures used to develop, test, deploy, and manage ML models reliably and efficiently. However, there is currently a lack of information on the practical implementation of MLOps in industrial enterprises.
    
    To address this issue, we conducted a multiple case study on MLOps in three large companies with dedicated MLOps teams, using established tools and well-defined model deployment processes in the Industry 4.0 environment. This study describes four of the companies' Industry 4.0 scenarios and provides relevant insights into their implementation and the challenges they faced in numerous projects. Further, we discuss MLOps processes, procedures, technologies, as well as contextual variations among companies. 
    
\end{abstract}


\section{Introduction}
\label{sec:introduction}

ML is increasingly used in Industry~4.0 enterprises~\cite{Angelopoulos2019Dec,DaXu2018Mar,ODonovan2018Jan}. 
MLOps is a set of techniques and tools for deploying machine learning models in production. Its focus is reducing delivery time, increasing automation, and ensuring reproducibility and repeatability. 
As more and more industrial ML applications are developed, the need for MLOps in this area is rising. 
There are few studies on how MLOps works in Industry~4.0. 
This raises the need to discuss the latest MLOps developments in industrial practice. 

This case study analyzes MLOps practices in three Industry~4.0 companies. 
These companies are all engaged in MLOps on a large scale and are located in Europe. 
We conducted three interviews with five employees to collect data about MLOps scenarios, organizational structures, roles, tools, and conditions in each company. 

Our findings reveal that MLOps practices vary significantly from one company to another. Each company has its unique approach to MLOps, depending on factors such as the industry, company size, and available resources. Further, the companies have different requirements, data, models, and production environments. 
Three trends are common across the companies. 
\textit{(I)} They had a dedicated team responsible for MLOps. \textit{(II)} They used a similar combination of open-source and commercial tools for MLOps. \textit{(III)} They had a well-defined process for deploying ML models into production. 

\section{Related Work}
\label{sec:related_work}

MLOps is a new and popular research topic. 
This section provides an overview of related work on MLOps in Industry 4.0. We focus on principles, tools, frameworks, challenges, organizational structures, and practical implementations. 
These works contribute to a comprehensive understanding of MLOps and its multifaceted challenges and applications. 
Despite the growing interest in MLOps, few case studies look at practical implementations of MLOps in different industries and environments~\cite{Mboweni2022Jul}. 
None of the identified case studies focuses on a broad range of MLOps practices concerning challenges, processes, and procedures and their variability in Industry 4.0. 
The body of research on MLOps encompasses various topics. These topics range from principles to practice, organizational challenges, technical challenges, and their relevance in Industry 4.0. 

\subsection{MLOps Principles, Tools, and Frameworks}
\label{sec:related_mlops}

Several papers address the foundational aspects of MLOps, discussing principles, tools, frameworks, and organizational structures that facilitate its implementation. The general idea behind MLOps is to utilize automation to bring machine learning models into production in an efficient manner, as well as to ensure quality improvement of these models both before and after deployment~\cite{Ruf2021Sep, Symeonidis2022Jan, Kreuzberger2023Mar}. 
In this context, MLOps tools, frameworks, and architectural considerations play an important role~\cite{Mboweni2022Jul,John2021Sep,Hewage2022Feb,Paleyes2023}.



\subsection{Organization and Socio-technical Challenges}
\label{sec:related_social_challenges}
Several studies shed light on the organizational and collaborative challenges in MLOps. 
Among these challenges are organizational aspects, such as differences in work culture, education, and multi-disciplinary collaboration~\cite{Kreuzberger2023Mar,Treveil2020Nov}. Specific challenges are, e.g., data organization, regulatory compliance, and data ownership. 

While some of these papers focus on challenges related to ML systems and organizational aspects~\cite{Kreuzberger2023Mar,Treveil2020Nov}, others identify organizational challenges, particularly in integrating collaborative MLOps practices~\cite{Granlund2021Sep} or deal with collaboration challenges, highlighting socio-technical aspects~\cite{Nahar2022May}. Further, socio-technical anti-patterns are described by~\cite{Mailach2023May}, drawing from a qualitative empirical study involving practitioners. 

\subsection{Development and Production Challenges}
\label{sec:related_tech_challenges}
In addition to organizational and socio-technical challenges, various articles address development and production challenges in MLOps. 
First, the concept of technical debt within machine learning systems is discussed~\cite{Sculley2015}. Further technical challenges and their implications for MLOps are highlighted by several authors~\cite{Banerjee2020,Symeonidis2022Jan,Garg2021Dec,Renggli2021Feb}. 
Sustainable challenges related to MLOps are explored by~\cite{Tamburri2020Sep}. Shankar et al.~\cite{Shankar2022Sep} provides insights into the experimental nature and technical challenges of MLOps. An analysis of challenges, particularly within the context of Industry~4.0, is provided in~\cite{faubel_mlops_industry}. 

\subsection{MLOps in Industry~4.0}
\label{sec:related_I40}
This section reviews relevant papers in the context of Industry~4.0. Shinha and Roy~\cite{Sinha2020May} comprehensively analyze smart factories and cyber-physical systems in Industry 4.0. They discuss technologies, management skills, architectures, and features required for such systems. The authors conducted case studies in two large companies to identify challenges and possible solutions. Additionally, they discuss the socio-economic impact of this development. Oluyisola et al.~\cite{Oluyisola2022Jan} propose a methodology for implementing MLOps in production planning and control, demonstrating its effectiveness within an industrial context. MLOps concepts, pipelines, life cycles, and tools that meet the specific needs for the IoT domain within Industry~4.0 are described and discussed by Bodor et al.~\cite{Bodor2023Mar}. Further, Schubert et al.~\cite{Schubert2023} develop a business-to-business integration framework for smart services in Industry~4.0, effectively integrating MLOps principles.

\section{Methods}
\label{sec:methods}

We conducted a multiple case study according to the guidelines on conducting and reporting case study research in software engineering by Runeson et al.~\cite{Runeson2009Apr,Runeson2012Feb}. 
The interviews were conducted online using a video conference platform. 
The companies with their focus, associate scenarios, interviewee roles, and interviewee IDs are summarized in Table~\ref{tab:case_companies}. In \textit{(a)}, the first column contains each token of the companies part of this study. The second column provides information about the company's focus. The third column refers to scenarios in Section~\ref{sec:mlops_scenarios}. The star * indicates that a company was part of a previous study~\cite{faubel_analysis_MLOps_practices}.
The table below, in \textit{(b)}, provides the interviewees' roles and assigned IDs in their respective companies.

\begin{table}
    \centering
    \caption{Characteristics of Case Companies and Interviewees.}
    \label{tab:case_companies}
    \vspace*{-1em}
    
    \begin{tabular}{llc}
        \toprule
        \textbf{Company} & \textbf{Focus} & \textbf{Scenario}\\
        \midrule
        A & Automotive manufacturer & 1 \\
        B* & Industrial devices and control systems & 2,4 \\
        C* & Digital technologies for industry & 1,3,4 \\
        \bottomrule
    \end{tabular}
 \\
 \textbf{(a) Case Companies} 
 \vspace{1em}
 
    \begin{tabular}{clc}
        \toprule
        \textbf{ID} & \textbf{Role} & \textbf{Company} \\
        \midrule
        IA1       & Software Architect & A  \\\hline
        IB1       & Project Manager & B     \\
        IB2        & Product Owner &        \\
        IB3        & Data Scientist &       \\\hline
        IC1       & Software Architect & C  \\
        IC2        & Data Scientist &       \\
        \bottomrule
    \end{tabular}
    \\
\textbf{(b) Interviewees}
\end{table}

\subsection{Research Questions}
The following research questions are addressed in this study to provide an overview of MLOps implementations in different companies:
\begin{description}
    \item[\textbf{RQ1:}]~What MLOps processes and procedures are used?
    \item[\textbf{RQ2:}]~What kind of solution structures (architectures) are used for MLOps?
    \item[\textbf{RQ3:}] To what extent do responses to RQ1 and RQ2 vary across and within companies?
\end{description}

Below, we describe the participants and how the method was instantiated in this study. 

\subsection{Participants}
\label{sec:participants}

To acquire the partners for this study, we contacted companies we know and with whom we have a corresponding trust relationship. We ultimately selected three companies that perform MLOps on a large scale. 

The following companies participated in this study: 


\textbf{Company A: }  
is a very large automotive manufacturer. The company develops ML models together with internal and external partners. For this purpose, they build their own MLOps platform. A platform architect was interviewed.

\textbf{Company B: } is a very large company building industrial devices and control systems for electrification, process automation, and robotics. This company develops ML systems using various MLOps platforms in different product groups and projects. Here, as opposed to the other cases, the focus is on developing ML solutions for their customers, not for themselves. A project manager, a product manager, and a data analyst were interviewed.

\textbf{Company C: }  is a large company that designs and manufactures electronics, software, and mechanics to provide digital technologies for industry. It develops ML systems to automate its production further and improve product quality. A software architect and a data analyst were interviewed.

\subsection{Data Collection} 
Semi-structured joint interviews were conducted following the guidelines of Boyce et al.\ \cite{boyce2006conducting} between August 2022 and February 2023. The interviewees were not selected by ourselves. Instead, we asked each company's contact person which employees were suitable for each interview question since they had an overview of their organizations. 
During the interview process, a structured guide was utilized, which was developed as explained in~\cite{boyce2006conducting}. The guide was designed to flow conversationally while still addressing all the research questions at hand and consists of 26 questions organized into 13 distinct steps~\cite{Runeson2012Feb}. Table~\ref{tab:interview_guide} summarizes these steps and the topic questions. 
\begin{table}
    \caption{Steps and Topics of the Interview Guide.}
    \label{tab:interview_guide}
    \vspace*{-1em}
    \begin{tabular}{cp{7cm}}
        \toprule
        \textbf{Steps} & \textbf{Topic}\\\toprule
         1 & Interviewees are invited via email and informed about the topic.\\
         2 & Introduction of the participants. Further, the interviewers introduce the topic MLOps, which is cross-checked by interviewees.\\
         3 & General questions about the use of Machine Learning.\\
         4 & Questions on organizational structure.\\
         5 & The interviewers ask about tools used in the context of the single MLOps activities.\\
         6 & Questions on conditions and legal requirements for tools.\\
         7 & Questions on data sets and requirements.\\
         8 & Questions on models and requirements.\\
         9 & Questions on communication methods and interfaces.\\
         10 & Questions on automation.\\
         11 & The interviewers ask about missing points and ideas.\\
         12 & Conclusion.\\
         13 & Final remarks.\\\bottomrule
    \end{tabular}
\end{table}

Each interview lasted approximately two hours. The interviewees were invited via email and informed about the topic, including subtopics for related questions. 

\subsection{Data Analysis}
\label{sec:data-analysis}
Qualitative data analysis was conducted according to the following steps~\cite{Runeson2009Apr}:
\begin{enumerate}
    \item Transcription.
    \item Creation of a coding scheme.
    \item Coding.
    \item Comparison of the information between different companies.
    \item Mapping codes to research questions.
    \item Drawing conclusions from the analysis.
\end{enumerate}

Two companies agreed to record and transcribe the session via professional transcription services. In another case, we had to take notes first and then create a transcript from the notes since no recordings were allowed. After transcription, the transcripts were coded using \textit{QDA Miner Lite}\footnote{https://provalisresearch.com/products/qualitative-data-analysis-software/freeware/}
, a qualitative data analysis software by Provalis Research. Coding was conducted according to~\cite[p.49]{guide_advanced_empirical_se}: During the coding process, a coding scheme was created by reading the interview transcripts and grouping passages into codes and categories. 
The initial list of codes and categories was based on the interview questions. Then, coding was performed iteratively. The transcripts were read multiple times. As they were re-read, new codes and categories were added that were appropriate for answering the research questions. 
After coding, the data from the different companies was compared and grouped, if applicable. Very little information was left out due to confidentiality reasons. This information included, for example, specific names of hardware components but no critical information to answer our research questions. Further, the codes were mapped to each research question. Finally, conclusions were drawn from the analysis. 


\subsection{Threats to Validity}
\label{sec:threats}
In this section, we will discuss threats to validity and propose mitigation strategies for them. To categorize these threats, we follow the guidelines suggested by Yin~\cite{yin2023Mar}  and Runeson~\cite{Runeson2009Apr} and divide them into three types, namely construct validity (Section \ref{sec:construct_validity}), internal validity (Section \ref{sec:internal validity}), external validity (Section \ref{sec:external validity}), and reliability (Section \ref{sec:reliability}). 
Further, we propose mitigation strategies for the threats in each category mentioned above. 

\subsubsection{Construct validity}
\label{sec:construct_validity}
First, construct validity refers to the extent of accurate observation. Here, we used the triangulation method. 
We tried interviewing two to four employees of the companies. Two interviewers asked questions and checked whether all points from the interview guidelines were addressed. As we could only interview one person in Company A, we recorded the transcript and had it checked. 

\subsubsection{Internal Validity}
\label{sec:internal validity}
Internal validity focuses on the degree of confidence that other factors or variables do not influence the causal relationship being tested. A case study involves an inference whenever an event cannot be directly observed. Based on evidence collected from the case study, investigators will infer that a particular event resulted from some earlier occurrence. We gathered participant feedback on our results to ensure correct inference and avoid misinterpreting codes and quotations. Further, incorrect groupings of codes and wrong results due to the previous threats were avoided by the other countermeasures and by conducting interviews with multiple researchers. 

\subsubsection{External validity}
\label{sec:external validity}
When conducting a case study, it is essential to consider whether the findings can be applied beyond the immediate study. This issue is related to analytic generalization. External validity defines the domain to which the study findings are applicable and generalizable. In this case study, the companies interviewed were from different Industry~4.0 domains and were large. The research questions were established during the research design phase to lay the foundation for addressing external validity. Various companies were interviewed to ensure the greatest possible external validity, given their size and domain. However, it should be noted that only three companies with mature practices were interviewed, which may affect generalizability. 

\subsubsection{Reliability}
\label{sec:reliability}
The reliability of the study findings is a crucial aspect that concerns the operability of the study design. It is essential that the study can be repeated with the same results to ensure reliability. Two researchers were involved in all data collection and analysis steps to minimize researcher bias. Additionally, the respective companies were included in the research process to provide feedback in different phases. During the interviews, structured questions were used to ensure consistency and transcripts of each case were created for further analysis. These measures ensure that the study findings are reliable and valid. 




\section{Findings and Discussion}
\label{sec:results}

\begin{figure*}[htbp]
    \includegraphics[width=1\textwidth,trim={20 18 20 18},clip]{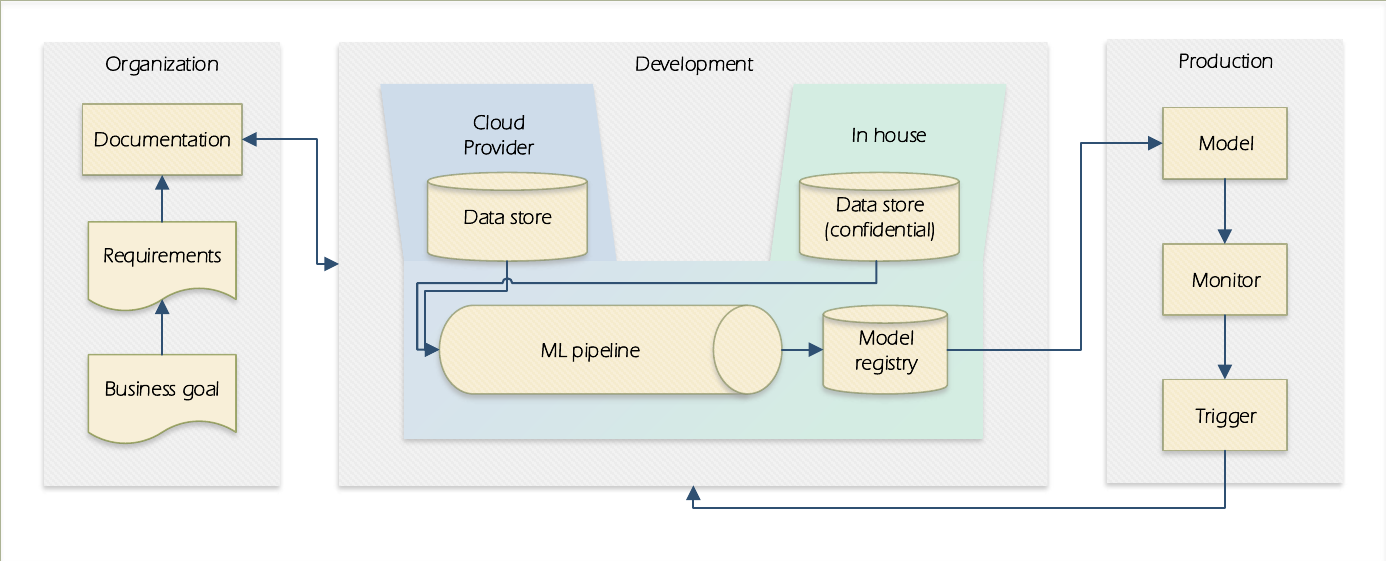}
    \caption{MLOps in Industry~4.0 overview.  \label{fig:summary}}
\end{figure*}

\outline{}{}{

- strukturieren: Stand der Technik oder in Zukunft, good practices, probleme

- was haben wir in den anderen Studien gesehen, wie verhalten sich unsere challenges dazu, z.B. concept drift

- Wie unterscheiden sich ausprägungen von MLOps

}

This section discusses practices, implementations, and the adoption of MLOps in the interviewed companies. 
At first, Section~\ref{sec:result_overview} gives an overview of the findings. 
Section~\ref{sec:result_challenges} describes the organization, development, and production challenges~(RQ1). 
Finally, software implementations are discussed in Section~\ref {sec:result_implementation}~(RQ2). 
Preliminary results and findings about roles, team structures, and tools in Company B and C were also reported in~\cite{faubel_analysis_MLOps_practices}. 


\subsection{Application Areas}
\label{sec:mlops_scenarios}
Based on our interviews, we identified four key MLOps scenarios. 


\textbf{Scenario 1} involves implementing a multi-purpose MLOps platform for various industrial automation needs. This platform considers hardware, infrastructure, and tool stacks needed for ML development and integration tests. With this platform, all types of training data can be ingested.

\textbf{Scenario 2} involves implementing MLOps solutions for predictive maintenance in process automation. This is crucial for ensuring stable operation in the process industry but requires customized solutions to account for variations in data depending on the customer and site. 

\textbf{Scenario 3} involves using MLOps solutions for automated visual inspection, which is increasingly used in a company due to a shortage of employees. Computer vision and neural networks detect defects in manufactured products using camera data.


\textbf{Scenario 4} involves anomaly detection to identify anomalies and deviations from normal behavior, e.g., in scenarios 2 and 3. This allows for monitoring and triggering of retraining.

\subsection{MLOps Overview}
\label{sec:result_overview}

To provide an overview of the structure of the findings, Figure~\ref{fig:summary} outlines the substructure and critical components in the machine learning projects, from organization and development to production. 

\subsubsection{Organization}
Organization refers to activities that ensure the project is planned, structured, and executed effectively. It is a critical aspect of a machine learning project and includes documentation, clear business goals, and requirements. Project managers are responsible for defining the business goals, which serve as a basis for development. They must ensure that the goals are aligned with the company's overall strategy and that they are measurable. Additionally, software architects define the software requirements based on the business goals, which specify what the solution should do, how it should do it, and what constraints it should operate under. By documenting these requirements, they ensure that the solution meets the business goals and is scalable and maintainable over time. Further, the requirements serve as a baseline for the development.

\subsubsection{Development}
After the business goals and requirements are defined, the next critical step is to design the software architecture. This is a crucial aspect of managing project quality and coherence, laying the foundation for the development process. Once the architecture is in place, data analysts and software engineers can develop a solution tailored to the project's requirements. Depending on the project's needs, this solution can run on a cloud-based or an in-house infrastructure managed by IT experts. Hybrid cloud infrastructures are also possible. The decision on the infrastructure usually depends on regulations, data confidentiality, or performance. Domain experts and process engineers familiar with industry processes are involved in developing ML solutions. They can assist with feature selection and evaluate solutions. If new solutions are to be tried, support roles or teams often assist in the development. The ML solution also requires data stores, pipelines, and model registries.

\subsubsection{Production}
In the deployment phase of a machine learning project, the ML models developed in the previous stages are deployed and applied in production, where they work with live data. The phases of monitoring and triggering were mentioned. Monitoring helps detect any issues with the model performance during production. Using the timestamps or performance thresholds, triggering allows for the necessary improvements.

As the companies differ significantly in this dimension, we discuss each of them in a separate subsection.

\subsection{Software Implementation}
\label{sec:result_implementation}

This chapter provides the findings regarding MLOps software implementations, tools, and methods.



\subsubsection{Development}
\label{sec:implementation_development}

From the interviews, we could derive the following insights on development. 

\textbf{Architecture:} 
MLOps architecture integrates various components, such as data management, data processing, model training, testing, deployment, and monitoring. 
Further, IA1 describes that the architecture for industrial MLOps consists of different layers: 
    \textcolor{blue}{\textit{``There are different layers. We[, the MLOps team,] are more in the tool stack layer. Below we have the infrastructure, [..] hardware and the execution environment.''} (IA1)}
\textit{(I)} The tool stack layer contains various programming frameworks, libraries, and tools. \textit{(II)} The infrastructure layer, on the other hand, provides container management and storage solutions. \textit{(III)}~The hardware layer covers the execution environment. 

These layers can also be found in Companies B and C, but interestingly, the team structure in Company A can mirror this architecture. For example, the MLOps team handles the tool stack layer, the container and storage teams manage the infrastructure layer, and the hardware team oversees the hardware layer. 

Regarding enterprise-specific architecture decisions, there are a few essential factors to consider. One of the key takeaways from our discussions is the importance of having clear guidelines in place for architecture. 
    \textcolor{blue}{\textit{``I think what's missing partially [..] is kind of good guideline, which path to follow. Because with machine learning, there's so many rabbit holes you could probably follow, and that lead to nowhere. And I think it's important to have a target first, and then follow a specific guide or track and then maybe get some assistance to find this path.''} (IB2)}
This helps ensure consistency across the organization and makes it easier for teams to work together. The companies prefer a structure with guidelines supported by frameworks and automated by software solutions. 



\textbf{Tools:}
The interviews highlight the use of various tools for infrastructure, data analysis, development, automation, monitoring, deployment, orchestration, and other purposes. These tools and their categories are summarized in Table~\ref{tab:tools}. 

\begin{table}
    \caption{List of tools used by the case companies.}
    \label{tab:tools}
    \vspace*{-1em}
        \begin{tabular}{ll}
            \toprule
            \textbf{Category} & \textbf{Tools} \\\toprule
            Infrastructure  & Kubernetes (3) SPARK (2) Ray (1)\\\hline
            Data Analysis   & Pandas (2) NiFi (1) ML workspace (1) \\
                            & Tensorflow (1) Keras (1) Pytorch (1) \\ 
                            & Grafana (1) Trino (1)\\\hline
            Development     &  Python (3)   Jupyter Notebook (1)\\
                            & Pycharm (1) Visual Studio (1)\\\hline 
            Automation      & Bamboo (2) Gitlab (2) Autoscaler (1) Argo (1) \\
                            & Atlassin Suite (1) Bitbucket (1) Teapot (1)\\
                            & Autokeras (1) Hyperopt (1) Optuna (1)\\\hline
            Monitoring      & splunk (1) Power BI (1) evidently (1)\\
                            & library detect (1) menelaus (1)\\\hline
            Deployment \&   & Seldon (3) MLflow (3) River (1) Kubeflow (1)\\ 
            Orchestration 
                            &  Openshift (1) Airflow (1) AWS (1) Kedro (1)\\
                            & Databricks (1) Polyaxon (1) Azure DevOps (1) \\\hline
            Other           & DVC (2) Ranger (1) Trino (1) Django (1) \\
                            & Elasticsearch (1) Jira (1) Osi PI (1) \\
                            & Postgres (1) mongoDB (1)\\\bottomrule
        \end{tabular}
\end{table}

\subsubsection*{Data and Model}

Different approaches are used within ML depending on a project's specific needs and goals. The following examples are identifiable in the interviews:

\textbf{Data Acquisition and Storage:} The interviews mentioned Various data acquisition and creation methods. For example, in Company C, an extensive data set was created by photographing wires in various ways: 
    \textcolor{blue}{\textit{``We removed the wire, took a picture, inserted it incorrectly, took a picture, also took pictures of good wirings, that's basically how we created the dataset. In the end the performance was really good.''} (IC1)} 
Such data sources require efficient mechanisms for data acquisition and storage. 
Interviewees mentioned various data storage and query technologies, such as \textit{Apache SPARK}, \textit{Apache NiFi}, and \textit{trino}. These technologies allow storing, managing, and querying large data sets from various sources and databases. 

\textbf{ML pipelines:}
One emerging trend is the move towards well-structured ML pipelines that simplify model training for non-experts. So, 
    \textcolor{blue}{\textit{``more and more use cases will go towards this well-prepared, well set-up pipeline that maybe even people without too much machine learning understanding can do.''} (IB1)}
This is a practical development, especially since customized solutions requiring extensive manual effort are still standard. However, achieving standardized structures in Industry 4.0 is challenging, given the need to balance standardization and customization to accommodate the varying needs of different industries. 


\subsubsection{Production}
Production is deploying, monitoring, and managing machine learning models and their infrastructure in the production environment. 

\textbf{Deployment:} For deployment, some companies use on-premise solutions, which can benefit performance and confidentiality. However, Platform-as-a-Service (PaaS) providers are becoming increasingly popular for enterprise applications. 
    \textcolor{blue}{\textit{``I think a lot really depends on the response time of the system you need. If it's milliseconds, if it's more seconds or if it's really minutes or even days, and if you would have a cloud solution that would run in milliseconds, probably we could move more to the cloud.''} (IB2)}
In one of the cases, we found that the company uses PaaS providers for most of its application landscape. However, there have yet to be any specific guidelines in place for particular use cases. There are various ways to deploy models, such as containerization with Docker, using Azure DevOps to automate deployments, and deploying models as web services in a Platform-as-a-Service environment. These approaches aim to improve efficiency and streamline the deployment process. As with all enterprise architecture decisions, it is essential to carefully consider the pros and cons of different approaches and choose the one that best meets the organization's needs now and in the future:
The decision to use cloud resources depends heavily on the specific business scenario. The model's response time is a factor to consider when making this determination. Further, considering resource requirements when selecting infrastructure is essential, as this can significantly impact costs. In certain situations, on-premise solutions may be more efficient for optimizing resources.

\textbf{Model Coordination:} Managing and coordinating models can be challenging, especially when deploying multiple models. 
    \textcolor{blue}{\textit{``When you build a model you might always have model performance problems. You have to retrain with other parameters or you have data drift or labels are missing or incorrect.''} (IA1)}
It is vital to consider experiment tracking, data management and administration as well. These factors are critical for ensuring the models are accurate and effective in achieving the desired outcomes. Further, 
    \textcolor{blue}{\textit{``The retraining-process should be included. We need to define thresholds in the monitoring and if it reaches the threshold a retraining should be triggered.''} (IA1)}
E.g., \textit{evidently} is used by the companies for the monitoring and to define automatic triggers for the retraining. However, automated retraining is still in its infancy in the companies. 

\section{Challenges}
\label{sec:result_challenges}


MLOps adoption in the case companies faces several challenges, from data management to deployment and explainability. While good practices are emerging, the challenges remain complex and multifaceted. Table~\ref{tab:challenges} summarizes the identified challenges. Several of these challenges were previously mentioned in the related work, while others are new. The socio-technical challenges are described in Section~\ref{sec:organization_challenges}, while the development challenges are described in Section~\ref{sec:development_challenges}, and the production challenges are described in Section~\ref{sec:production_challenges}. 

\begin{table}[htbp]
    \caption{Challenges.\label{tab:challenges}}
    \vspace*{-1em}
    \begin{center}
        \begin{tabular}{lll}
        \toprule
            \multicolumn{3}{c}{\textbf{Organization} }\\\toprule
            \textbf{Challenge Context} & \textbf{Interviews} & \textbf{Literature} \\
            \toprule
            Human-machine interaction & B, C & \\\toprule
            \multicolumn{3}{c}{\textbf{Development} }\\\toprule
            Automation and deployment & A, B, C & \cite{Kreuzberger2023Mar,Garg2021Dec,faubel_mlops_industry}\\
            Standardized data access & B, C & \\
            Variety vs. Architecture & A, B, C & \\\toprule
            
            \multicolumn{3}{c}{\textbf{Production} }\\\toprule
            Model and explainability & B, C & \\
            Missing production metrics & B & \cite{Mailach2023May,Symeonidis2022Jan,Garg2021Dec,Renggli2021Feb}\\
            and triggers & & \\
            \bottomrule
        \end{tabular}
    \end{center}
\end{table}

\subsection{Socio-Technical}
\label{sec:organization_challenges}
This section discusses socio-technical challenges that refer to the issues that arise when implementing new technologies in a social or organizational context. 

\textbf{Human-machine interaction:} 
Human involvement in high-risk scenarios brings potential challenges in defining the boundary between automated and human decision-making. For example, Company B states that automated decisions are not always desired, e.g., if a human made a change:
    \textcolor{blue}{\textit{``It could really be that the operator changed something physically in the plant, and then of course the model won't fit anymore[..]''}~(IB2)}
In a case where a human has caused a state change that results in a loss of performance of the ML model, automated retraining without involving humans can be problematic. 
Finding the right balance between automated decisions and human judgment, especially in critical situations, requires careful planning and clear guidelines. 

\subsection{Development}
\label{sec:development_challenges}

We have identified developmental challenges w.r.t.\ the MLOps \textit{architecture} \textit{automation and deployment}, and \textit{standardization and data access}.

\textbf{Variety vs. Architecture:} Experts share optimism about establishing software architectures for ML development and deployment. 
The consensus is that while a standardized reference architecture with minor variations is feasible, it's vital to strike a balance between standardization and flexibility. 
    \textcolor{blue}{\textit{``I think on the high level, I would be optimistic that there can be one MLOps architecture or several ones. And there's probably, I mean, architectures are taking decisions about alternatives, right? [..] But, the more we go into the details, of course it will vary.''} (IB1)}
Different industries have unique approaches to MLOps. The architectural challenges they face can be influenced by the industrial application and the specific systems. 

\textbf{Automation and Deployment:} As companies strive towards greater automation in their development processes, it becomes increasingly important for them to have reliable CI (continuous integration)/CD (continuous deployment) tracking and proper pipelines of models. This is already predominantly ensured by existing tools. However, the challenge of manual model engineering persists, even as companies explore using automated ML tools. While these tools hold great potential, they have yet to reach a level of sophistication that can completely replace the expertise of human engineers. 
For example, there are still problems in finding all the properties leading to a performance change: 
    \textcolor{blue}{\textit{``The other thing is more this kind of creeping problems, that the plant is aging or that the valves are basically no more as linear as they used to be [..] this is a challenge in any respect, because then a lot of things may change and you never know if it's really a problem of the model itself, or it's more a problem of something of the underlying data [..].''} (IB2)}
These changes in the underlying data, some of which are human-made or due to technical reasons, e.g., failures, still need to be detected by humans and cannot be easily automated. 

\textbf{Standardization and Data Access:} Deploying systems and processes, particularly in diverse industrial settings, can be challenging to standardize. Even though there is a desire for standardized procedures, the wide range of differences in shop floors and environments makes it difficult to overcome. 
Utilizing standardized data access platforms can provide advantages. These platforms offer extensive support, and the process of accessing vital data is simplified.
    \textcolor{blue}{\textit{``[..] the platforms often give access to data. In a standardized way, that is a very big support and making life much, much easier.''} (IB1)}
However, an interviewee has observed that these data platforms need more and better labeling capabilities, which affects the meaningful and efficient organization and categorization of data for ML purposes (IC1). Further, it is important to note that commercial data platforms may lead to vendor lock-in, limiting flexibility in switching to alternative platforms if needed. For this reason, custom database systems are created by individual companies for labeling purposes, although this may require a significant investment of time. 

\subsection{Production}
\label{sec:production_challenges}

Here, challenges in production concerning models, explainability, and metrics are described. 

\textbf{Model and Explainability:} Managing and coordinating models can be challenging, especially when deploying multiple models and explanation methods. To ensure good model performance, it is also essential to consider data management, administration, and monitoring for both models and explainers. These factors are critical for ensuring the models are accurate and effective in achieving the desired outcomes. The interviewed companies commonly use model monitoring, whereas explainability solutions are less prevalent. One example is the power plant console: 
    \textcolor{blue}{\textit{``So basically, what you get on this console is a representation of the physical plant and you get the measurements of the physical plant and that's to it, and it's all up to the human in front of it to do the interpretation.''} (IB1)}
The importance of model documentation, particularly for explainability, is rising due to the need for trustworthiness, maintainability, and traceability in ML. This is one of the aspects where explanations are needed. Further, there is a functional need for manual feedback on the explanations to provide the ability to change the representation, show different kinds of explanations, or report inadequate explanations for maintenance reasons. 

\textbf{Missing Production Metrics:} Data and model monitoring in production can be challenging since companies do not have accurate metrics to detect malfunctions or to trigger an ML model retraining: 
    \textcolor{blue}{\textit{``[..] for monitoring, we do it at a very basic level [..] in monitoring we could also include tools for explainability [..] and maybe even triggers that for example, retrain a model once certain metric reaches some value [..] we [..] Haven't developed at that much [..]''} (IC2)} 

\subsection{Trends and Future Developments}
\label{sec:development}
Some trends and future developments emerge as focus areas for further development in the interviews. 

\textbf{Enhancing Explainability:} The black-box nature of some machine learning models has raised concerns about their need for explainability. 
    \textcolor{blue}{\textit{``The main thing is how can we make our black box models [..] more explainable so we can actually show them why the model made such a decision and if it actually was correct or not.''} (IC1)}
Future developments are geared towards making models more explainable during training and inference. Researchers are exploring methods to provide insights into why a model makes specific decisions, promoting trust and enabling humans to validate and understand the outcomes.



\textbf{Deployment:}
The interview with one of the companies indicates that containerization is expected to become prevalent within the next few years. Company C is already using container deployment for all its solutions. Company B is 
    \textcolor{blue}{\textit{``[..] moving there as well. I mean, that is a time cycle where we say: that's going to be the future. So, maybe [..], it makes sense to jump to this assumption that there will be a way to deploy things in a containerized fashion.''} (IB1)}
The transition to Docker-based deployment has challenges. Some issues include the run time performance of deploying containers on older edge devices.

\section{Discussion}
\label{sec:results_future_directions}

Based on interviews, this paper describes MLOps scenarios, organization, development, and production aspects in ML projects, implementations, tools, methods, and challenges. 
Although MLOps is a prominent research topic, only a few case studies of specific MLOps solutions exist. To fill this gap, we interviewed three companies in the Industry 4.0 domain that have mature MLOps platforms in place. In this study, we provide valuable insights into the current MLOps practices in Industry 4.0. The findings suggest that companies should develop their unique approach to MLOps, depending on their specific situation. However, companies can learn from each other's experiences in these specific situations to improve their MLOps processes. 
Although diverse goals are pursued with MLOps, there are clear overlaps in the procedures, roles, and responsibilities in the respective companies during implementation.

The novel challenges that we did not find covered by the literature are in the domain of human-machine interaction, standardized data access, variety vs. architecture, and model and explainability. 
Further, we identified challenges in the area of knowledge sharing and documentation, automation and deployment, and missing production metrics and triggers, which are already covered by the literature. 
These challenges should be addressed in the future.

\section{Conclusion}
\label{sec:conclusion}

In conclusion, MLOps processes and roles are similar among the companies. 
Still, the operational part has a need for explainability, automated monitoring, and retraining. We identify standards and show that the fundamental technology does not differ significantly from general MLOps technologies. However, structural considerations are necessary in exceptional cases, limiting the tool selection. Further, the integration of diverse data sources remains difficult. Cross-domain collaboration between domain experts and ML experts can help build solutions to address these industry challenges. 


This information will be used in future investigations to build an MLOps platform that adapts to different industrial scenarios. We also want to encourage further interviews and case studies to extend and support our findings and investigate other specific MLOps cases.

\section*{Acknowledgment}

This work is supported by the project EXPLAIN, funded by the German Federal Ministry of Education under grant 01—S22030E. Any opinions expressed herein are solely by the authors and not the funding agency. 






\renewcommand*{\bibfont}{\footnotesize}
\printbibliography


\end{document}

%% file: aux_commands.tex
\usepackage{ifthen} 
\usepackage{csquotes}

\newboolean{auxinfo}
\setboolean{auxinfo}{true}   

\newboolean{showoutline}
\setboolean{showoutline}{false}

\newboolean{showcomment}

\setboolean{showcomment}{true} 

\usepackage{ifthen}

\ifthenelse{\boolean{auxinfo}}
  {}
  {
    \setboolean{showoutline}{false}
    \setboolean{showcomment}{false}
  }

\usepackage{savesym}
\usepackage{xcolor}
\savesymbol{Bbbk}
\usepackage{amssymb}
\restoresymbol{AMS}{Bbbk} 

\ifthenelse{\boolean{showcomment}}
   {\newcommand{\nb}[2]{\fcolorbox{gray}{yellow}{\bfseries\sffamily\scriptsize#1}{\sf\small$\blacktriangleright${\em #2}$\blacktriangleleft$}}
   \newcommand{\working}[1]{\fcolorbox{gray}{yellow}{{\bf #1}\emph{\scriptsize---in progress---}}}
   \newcommand{\TBD}[1]{\fcolorbox{gray}{yellow}{{\bf #1}\textbf{TBD}}} 
  }
  {\newcommand{\nb}[2]{}{}
   \newcommand{\working}[1]{}
   \newcommand{\TBD}[1]{} 
  }

\usepackage{todonotes}
\ifthenelse{\boolean{showoutline}}{
	\newcommand{\outline}[3]{
		~\newline 
		\fcolorbox{red}{white}{
			\parbox{\columnwidth-13.465pt}{
				\ifthenelse{\equal{#1}{}}{
					\ifthenelse{\equal{#2}{}}{
						\noindent\colorbox[rgb]{0.65,0.16,0}{\textcolor[rgb]{1,1,1}{\textbf{Outline}}}
					}{
						\colorbox[rgb]{0.65,0.16,0}{\textcolor[rgb]{1,1,1}{\textbf{Outline -- Responsible: #2}}}
					}
				}{
					\ifthenelse{\equal{#2}{}}{
						\noindent\colorbox[rgb]{0.65,0.16,0}{\textcolor[rgb]{1,1,1}{\textbf{#1 page(s)}}}
					}{
						\colorbox[rgb]{0.65,0.16,0}{\textcolor[rgb]{1,1,1}{\textbf{#1 page(s) -- Responsible: #2}}}
					}
				}
				#3
			}
		}
	}
}{
	\newcommand{\outline}[3]{}
}

\newcommand\defauxcomm[1]{
       \expandafter\newcommand\csname #1NB\endcsname[1]{\nb{#1}{##1}}
       \expandafter\newcommand\csname #1WK\endcsname{\working{#1}}
       \expandafter\newcommand\csname #1TBD\endcsname{\nb{TBD #1}}
    } 

\defauxcomm{ks} 
\defauxcomm{he} 
\defauxcomm{se} 
\defauxcomm{ck} 
\defauxcomm{lg} 
\defauxcomm{ak} 
\defauxcomm{lf} 
\defauxcomm{kl} 
\defauxcomm{cs} 
\defauxcomm{aa} 

\defauxcomm{mn}
\defauxcomm{fk}
\defauxcomm{ek}

\usepackage[normalem]{ulem}
\ifthenelse{\boolean{showcomment}}
    {\newcommand{\strike}[1]{\textcolor{red}{\sout{#1}}}}
    {\newcommand{\strike}[1]{}}